\documentclass[12pt]{iopart}
\usepackage{epsf}
\usepackage{graphicx}

\begin{document}
\title{Richtmyer - Meshkov instability in a spherical target with density variation}
\author{Labakanta Mandal $^{*}$, Sourav Roy, Rahul Banerjee, Manoranjan Khan and M.R.Gupta}
\address{Department of Instrumentation Science \& Center for Plasma Studies\\ Jadavpur University, Kolkata-700032, India}
\ead{labakanta@gmail.com}
\date{}

\begin{center}
\line(1,0){500}
\end{center}
\begin{abstract}

The motion of unstable fluid interface due to Richtmyer - Meshkov
(RM) instability incorporating with density variation has been
studied in a spherical target using Lagrangian formulation. During
the compression in Inertial Confinement Fusion (ICF) process, the
density of deuterium - tritium (DT) fuel increases 1000 times
greater than the density of gaseous DT fuel within the core of
spherical target. We have extended the feature of density
variation [PRA,84-Mikaelian \& Lindl] in spherical geometry. Due
to convergent shock impingement, the perturbed interface will be
nonspherical which leads to the density variation in both radial
as well as in polar angle. We have shown that the interface of
perturbed surface decreases with time to reach a minimum and then
kick back to gradual increase. As the perturbed radius decreases,
the density increases and reaches a maxima corresponding to a
minima of perturbed radius. This is the practical situation of
density characteristics during implosion of ICF. The numerical
results based on our analytical work show a good qualitative
agreement with some experimental as well as simulation results.

\end{abstract}
\begin{center}
\line(1,0){500}
\end{center}



\section{Introduction}

The direct driven Inertial Confinement Fusion (ICF) process can be
divided into four stages, i.e., the laser radiation, the
implosion-compression, the fusion ignition and the
deuterium-tritium (DT) fuel burn. Mainly, in the second stage, the
hydrodynamical interfacial fluid instabilities can act. The
Rayleigh-Taylor (RT) instability can occur during the two phases
of the ablative acceleration and the compression deceleration of
the ICF implosion. In addition to RT instability, RM instability
can occur when shocks pass through the irregular fluid interface.
The nonspherical implosion in ICF process generates the shear flow
causing the Kelvin-Helmholtz (KH) instability. {\bf The growth of
these instability can be affected by the density ratio of two
fluids, geometry of the interface, compressibility effects, heat
conduction and mass flow at the interface}. These instabilities
should be mitigated for efficient compression in ICF experiment.

Mikaelian and Lindl${\cite{km84}}$ explained that the smooth
density gradient at the interface stabilize the hydrodynamic
instability in ICF situation. However, they considered a constant
and exponential density gradient at the interface. Initially this
work started by Plesset ${\cite{pl54}}$ for stability condition of
underwater explosion considering the velocity along the radial
direction only. He discussed the stability and instability
conditions depending on the pressure difference at the interface
of bubble. Growth of perturbed interface due to density
${\cite{km83A}}-{\cite{km86A}}$ variation has been discussed in
different geometry ${\cite{vg05}}$. RT growth stabilization effect
has been showed very well by hot spot dynamics
${\cite{as07}}-{\cite{jg05}}$ in ablation situation of ICF.
However, Ramshaw ${\cite{jr99}}$ extended
Mikaelian's${\cite{km90l}},{\cite{km90}}$ work using Lagrangian
formulation in hydrodynamical model in a spherical geometry. In
this context, several authors $\cite{js74}-\cite{rp96}$ pointed
out that during implosion(acceleration phase) the fluid is
compressed hence the fluid density increases rapidly inside the
sphere. Similarly, during the blow off(deceleration phase) the
fluid is relaxed to expand and fluid density falls exponentially
from the perturbation surface. During robustness compression
 ${\cite{xr09}}-{\cite{sa11}}$
of direct driven ICF target, the density can be compressed by 100
to 200 times of liquid DT density at cryogenic temperature
${\cite{mc88n}}-{\cite{mc88p}}$. Perturbation amplitudes and its
growth rates of hydrodynamical instabilities of laser fusion
targets has been studied in acceleration and deceleration phase
including the heat conduction effect
${\cite{sa86}}-{\cite{tm06}}$using Lagrangian formulation.
Numerical,simulation and experimental
results${\cite{qz98}}-{\cite{jz08}}$ of RMI are performed in
cylindrical geometry considering implosion and explosion
situation. However, the growth of RM instability can be suppressed
due to different shape of the interface
${\cite{km05f}}-{\cite{mrg07}}$. Recently, X-ray and proton
radiography technique has been used to study the fast ignition
implosion in cylindrical geometry ${\cite{lv11}}-{\cite{lv11pl}}$.
The simulation result well agree with the experimental result. So,
density variation and sphericity are two key factor to understand
clearly for successful of ICF process.

In this paper, we investigate the nonlinear evolution of perturbed
interface in spherical geometry considering mean interface
$[R(t)]$ and angular amplitude both are dynamical variables. As a
result, the growth and growth rate of nonspherical interface due
to implosion has been changed for both variable with time. Hence,
the nonspherical irregular interface changes explicitly with
radius, polar angle and time. Analytically and numerically we have
shown the density variation for compression-blow off situation of
ICF. Theoretically high compression may be achieved depending on
the density variation at the interface of two fluids. Polar plot
also shows the time evolution of the interface in spherical
geometry. Growth reduction has been shown for high value of mode
number for a particular initial conditions.

The paper is organized as follows: section 2, deals with basic
assumptions and equation of continuity. Expressions of kinetic
energy of two fluids followed by Lagrangian equation of motion
have been discussed in section 3. The nonlinear equations which
describe the growth and the growth rate of perturbed interface are
presented in section 4. Section 5 contains results and
discussions. Finally, we conclude the results with a hope of
future study on this work.

\section{Basic equation}
We have considered two concentric spherical shells having two
incompressible, immiscible and inviscid fluids of density $\rho_1$
and $\rho_2$. The unperturbed interface at any time is given by
$r=R(t)$. The lighter fluid is bounded in the inner shell at
$R_1<\hat r<R(t)$ and the heavier fluid is ceiling at radius
$R_2$, where, $R_1\ll R(t)\ll R_2$, as shown in Fig.1. Here we
assume that the interface is perturbed to $r=\hat r$ which is
given by
\begin{eqnarray}\label{eq:1}
\nonumber
\hat{r}(t,\theta)=\hat{R}(t)+\sqrt{\frac{2l+1}{2}}a_l(t)P_l(cos\theta)\\
\hat{r}(t,\theta)=\Big[\hat{R}(t)+\delta
R_{a_l}\Big]+\sqrt{\frac{2l+1}{2}}a_l(t)P_l(cos\theta)
\end{eqnarray}
where, $a_l\ll R$ and $\hat R, a_l, P_{l}(cos\theta)$ are mean
radius of perturbed interface, angular displacement and Legendre
polynomial of mode number ${\it l}$ respectively  and $\delta
R_{a_l}=R[1-(\frac{a_l}{R})^{2}]$. Here we ignore the azimuthal
dependence of perturbed interface due to spherical symmetry.
However, simulation $\cite{hs90}$ study shows that there is a weak
dependence of growth rate on azimuthal angle.

Now we assume that the fluid motion is irrotational in the
interface region,i.e., its obey Laplace's equation $\nabla ^{2}
\Phi=0$. So the velocity potentials for lighter and heavier fluids
are
\begin{eqnarray}\label{eq:2}
\Phi_1=-\frac{R^{2}\dot{R}}{r}+\frac{1}{\it l}\sqrt{\frac{2 {\it
l}+1}{2}}\left(R\dot{a_l}+2\dot{R}a_l\right)\left(\frac{r}{R}\right)^{\it
l}P_{l}(cos \theta)
\end{eqnarray}
\begin{eqnarray}\label{eq:3}
\Phi_2=-\frac{R^{2}\dot{R}}{r}-\frac{1}{{\it l}+1}\sqrt{\frac{2
{{\it
l}+1}}{2}}\left(R\dot{a_l}+2\dot{R}a_l\right)\left(\frac{R}{r}\right)^{{\it
l}+1}P_{l}(cos \theta)
\end{eqnarray}
respectively.

Due to shock impingement, the nonspherical perturbed interface
leads to the density variation as follows
\begin{eqnarray}\label{eq:4}
\rho_{1}(r,\theta,t)=\rho_1+\rho_{11}(r,t)P_l(cos \theta);
\qquad\qquad R_1< r < \hat r
\end{eqnarray}
\begin{eqnarray}\label{eq:5}
\rho_{2}(r,\theta,t)=\rho_2+\rho_{21}(r,t)P_l(cos \theta);
\qquad\qquad \hat r< r \le R_2
\end{eqnarray}
for lighter and heavier fluid, respectively. This type of density
variation feature was not considered in previous studies.

\subsection{Equation of continuity}

The equation of continuity in spherical polar coordinate can be
written as
\begin{eqnarray}\label{eq:6}
\frac{\partial \rho}{\partial t}+u_{1r}\frac{\partial
\rho}{\partial r}+\frac{u_{1\theta}}{r}\frac{\partial
\rho}{\partial t}=0
\end{eqnarray}
where, $u_{1r}=\frac{\partial \Phi}{\partial r}$ and
$u_{1\theta}=\frac{1}{r}\frac{\partial \Phi}{\partial \theta}$ are
radial and transverse component of fluid velocity, respectively.
Substituting the value of $u_{1r},u_{1\theta}$and $\rho_1$, we get
the following equation for lighter fluid
\begin{eqnarray}\label{eq:7}
\nonumber \frac{\partial \rho_{11}}{\partial
t}+\left[\frac{R^2\dot R}{ r^2}+\sqrt{\frac{2{\it
l}+1}{2}}\left(R\dot{a_l}+2\dot{R}a_l\right)\left(\frac{r^{({\it
l}-1)}}{R^{\it l}}\right)P_{l}(cos \theta)\right]\frac{\partial
\rho_{11}}{\partial r}\\+\frac{1}{\it l}\sqrt{\frac{2 {\it
l}+1}{2}}\left(R\dot{a_l}+2\dot{R}a_l\right)\left(\frac{r^{({\it
l}-1)}}{R^{\it l}}\right)\frac{d P_{l}}{d\theta}(cos \theta)=0
\end{eqnarray}
Neglecting the product term $(a_l\times \rho_{11})$, we get the
following partial differential equation
\begin{eqnarray}\label{eq:8}
\frac{\partial \rho_{11}}{\partial t}+\frac{R^{2}\dot
{R}}{r^{2}}\frac{\partial \rho_{11}}{\partial r}=0
\end{eqnarray}
The solution of the above p.d.e is given by,
\begin{eqnarray}\label{eq:9}
\rho_{11}(r,t)=\rho_{11}e^{\frac{\mu}{3}\big[R^{3}(t)-r^{3}\big]}
\end{eqnarray}
Similarly, we get the density variation for heavier fluid as
follows
\begin{eqnarray}\label{eq:10}
\rho_{21}(r,t)=\rho_{21}e^{\frac{\mu}{3}\big[R^{3}(t)-r^{3}\big]}
\end{eqnarray}
where, $\mu$ is a positive constant for implosion situation.\\ The
mass conservation at the interface gives us the following relation
\begin{eqnarray}\label{eq:11}
\frac{\rho_{11}}{\rho_1}=\frac{\rho_{21}}{\rho_2}
\end{eqnarray}
\section{Kinetic energy}

To study the equation of motion of perturbed interface we have
calculated the kinetic energy of the two fluids.

\subsection{Kinetic energy of lighter fluid}
The kinetic energy of lighter fluid can be calculated in the
following way
\begin{eqnarray}\label{eq:12}
\nonumber T_1=\frac{1}{2}\int d\Omega\int ^{\hat r}
_{R_1}\rho_1(r,\theta,t)\Bigg[|\nabla \Phi_1|^{2}\Bigg]r^{2}dr\\
\nonumber \hskip 15pt =\frac{1}{2}\int d\Omega\int ^{\hat r}
_{R_1}\Bigg[\rho_1+\rho_{11}(r,t)P_l(cos\theta)\Bigg]\left[\Bigg(\frac{\partial\Phi_1}{\partial
r} \Bigg)^{2}+\frac{1}{r^2}\Bigg(\frac{\partial \Phi_1}{\partial
\theta}\Bigg)^2\right]r^{2}dr\\
\fl \nonumber
=\frac{1}{2}\int d\Omega\int ^{\hat r}
_{R_1}\Bigg[\rho_1+\rho_{11}(r,t)P_l(cos\theta)\Bigg]\Bigg[\frac{R^4\dot
R^2}{ r^4}+2\frac{R^2\dot R}{r^2}\sqrt{\frac{2{\it
l}+1}{2}}\Bigg(R\dot{a_l}+2\dot{R}a_l\Bigg)\left(\frac{r^{({\it
l}-1)}}{R^{\it l}}\right)P_{l}(cos\theta)\\
\fl \nonumber \hskip 2pt +\Bigg(\frac{2l+1}{2}\Bigg)\Bigg(R\dot
{a_l}+2\dot Ra_l\Bigg)^2\Bigg(\frac{r^{2({\it l}-1)}}{R^{2{{\it
l}}}}\Bigg)\Bigg(P_{\it l}(cos\theta)\Bigg)^{2}+\frac{1}{{\it
l}^2}\Bigg(\frac{2{\it l}+1}{2}\Bigg)\Bigg(R\dot {a_l}+2\dot
Ra_l\Bigg)^2\Bigg(\frac{r^{2({\it l}-1)}}{R^{2{{\it
l}}}}\Bigg)\Bigg(\frac{dP_{\it
l}}{d\theta}\Bigg)^{2}\Bigg]r^{2}dr\\
\fl=T_{10}+T_{1r\theta}
\end{eqnarray}
Where, $T_{10}$ and $T_{1r\theta}$ are the kinetic energy coming
from homogeneous density $\rho_1$ and inhomogeneous density
$\rho_{11}(r,t)$ part of the lighter fluid, respectively. Now
using Eq. 9, we get $T_{1r\theta}$ as follows,
\begin{eqnarray}\label{eq:13}
\fl \nonumber T_{1r\theta}=\frac{\rho_{11}}{2}\int
P_{l}(cos\theta) d\Omega\int ^{\hat r} _{R_1}\Bigg[\frac{R^4\dot
R^2}{ r^2}+2{R^2\dot R}\sqrt{\frac{2{\it
l}+1}{2}}\Bigg(R\dot{a_l}+2\dot{R}a_l\Bigg)\left(\frac{r^{({\it
l}-1)}}{R^{\it l}}\right)P_{l}(cos\theta)\\
\fl \nonumber \hskip 100pt +\Bigg(\frac{2l+1}{2}\Bigg)\Bigg(R\dot
{a_l}+2\dot Ra_l\Bigg)^2\Bigg(\frac{r}{R}\Bigg)^{2 {\it
l}}\Bigg(P_{\it l}(cos\theta)\Bigg)^{2}\\
\fl \hskip 120pt +\frac{1}{{\it l}^2}\Bigg(\frac{2{\it
l}+1}{2}\Bigg)\Bigg(R\dot {a_l}+2\dot
Ra_l\Bigg)^2\Bigg(\frac{r}{R}\Bigg)^{2 {\it l}}\Bigg(\frac{dP_{\it
l}}{d\theta}\Bigg)^{2}\Bigg]e^{\frac{\mu}{3}\Big[R^{3}(t)-r^{3}\Big]}dr
\end{eqnarray}
Similarly, we have calculated the kinetic energy for heavier fluid
in the preceding section.
\subsection{Kinetic energy of heavier fluid}
\begin{eqnarray}\label{eq:14}
\nonumber T_2=\frac{1}{2}\int d\Omega\int ^{R_2}_{\hat r}
\rho_2(r,\theta,t)\Bigg[|\nabla \Phi_2|^{2}\Bigg]r^{2}dr\\
\nonumber \hskip 15pt =\frac{1}{2}\int d\Omega\int ^{R_2}_{\hat
r}\Bigg[\rho_2+\rho_{21}(r,t)P_l(cos\theta)\Bigg]\left[\Bigg(\frac{\partial\Phi_2}{\partial
r} \Bigg)^{2}+\frac{1}{r^2}\Bigg(\frac{\partial \Phi_2}{\partial
\theta}\Bigg)^2\right]r^{2}dr\\
\fl \nonumber =\frac{1}{2}\int d\Omega\int ^{R_2}_{\hat r}
\Bigg[\rho_2+\rho_{21}(r,t)P_l(cos\theta)\Bigg]\Bigg[\frac{R^4\dot
R^2}{r^4}+2\frac{R^2\dot R}{r^2}\sqrt{\frac{2{\it
l}+1}{2}}\Bigg(R\dot{a_l}+2\dot{R}a_l\Bigg)\left(\frac{R^{{(\it
l}+1)}}{r^{({\it
l}+2)}}\right)P_{l}(cos\theta)\\
\fl \nonumber \hskip 200pt +\Bigg(\frac{2l+1}{2}\Bigg)\Bigg(R\dot
{a_l}+2\dot Ra_l\Bigg)^2\Bigg(\frac{R^{2{({\it l}+1)}}}{r^{2({\it
l}+2)}}\Bigg)\Bigg(P_{\it l}(cos\theta)\Bigg)^{2}\\
\nonumber \hskip 80pt+\frac{1}{({\it l}+1)^2}\Bigg(\frac{2{\it
l}+1}{2}\Bigg)\Bigg(R\dot {a_l}+2\dot
Ra_l\Bigg)^2\Bigg(\frac{R^{2{({\it l}+1)}}}{r^{2({\it
l}+2)}}\Bigg)\Bigg(\frac{dP_{\it
l}}{d\theta}\Bigg)^{2}\Bigg]r^{2}dr\\
\fl=T_{20}+T_{2r\theta}
\end{eqnarray}
Where, $T_{20}$ and $T_{2r\theta}$ are the kinetic energy coming
from homogeneous density $\rho_2$ and inhomogeneous density
$\rho_{21}(r,t)$ part of the lighter fluid, respectively, and
$\int d\Omega=2\pi\int \sin{\theta}d\theta$=solid angle. Now using
Eq. 10, we get $T_{2r\theta}$ as follows,
\begin{eqnarray}\label{eq:15}
\fl \nonumber T_{2r\theta}=\frac{\rho_{21}}{2}\int d\Omega\int
^{R_2}_{\hat r} \Bigg[\frac{R^4\dot R^2}{r^4}+2\frac{R^2\dot
R}{r^2}\sqrt{\frac{2{\it
l}+1}{2}}\Bigg(R\dot{a_l}+2\dot{R}a_l\Bigg)\left(\frac{R^{{(\it
l}+1)}}{r^{({\it
l}+2)}}\right)P_{l}(cos\theta)\\
\fl \nonumber \hskip 100pt +\Bigg(\frac{2l+1}{2}\Bigg)\Bigg(R\dot
{a_l}+2\dot Ra_l\Bigg)^2\Bigg(\frac{R^{2{({\it l}+1)}}}{r^{2({\it
l}+2)}}\Bigg)\Bigg(P_{\it l}(cos\theta)\Bigg)^{2}\\
\fl \hskip 50pt+\frac{1}{({\it l}+1)^2}\Bigg(\frac{2{\it
l}+1}{2}\Bigg)\Bigg(R\dot {a_l}+2\dot
Ra_l\Bigg)^2\Bigg(\frac{R^{2{({\it l}+1)}}}{r^{2({\it
l}+2)}}\Bigg)\Bigg(\frac{dP_{\it
l}}{d\theta}\Bigg)^{2}\Bigg]e^{\frac{\mu}{3}\big[R^{3}(t)-r^{3}\big]}r^{2}dr
\end{eqnarray}
\subsection{Lagrangian equation of motion}

Now we have formulated the problem using Lagrangian equation of
motion. The present problem is related to the RM instability i.e.,
shock impingement at the interface of two fluid and hence
neglected the potential energy. So, we have Lagrangian equation of
motion as follows
\begin{eqnarray}\label{eq:16}
\frac{d}{dt}\Bigg(\frac{\partial T}{\partial
\dot{q}}\Bigg)-\frac{\partial T}{\partial q}=0
\end{eqnarray}
where, $T=T_{1r\theta}+T_{2r\theta}$ and $q \rightarrow$ R and
$a_l$ both are dynamical vriable.
\section{Nonlinear equation}
Using eqs.13 \& 15, We get from eq. 16, a pair of coupled
nonlinear second order following differential equation. The
detailed calculations are given in appendix.
\begin{eqnarray}\label{eq:17}
\fl
f_{1R}(R,a_l)\ddot{R}+f_{2R}(R,a_l)\ddot{a_l}+f_{3R}(R,a_l)\dot{R^2}+f_{4R}(R,a_l)\dot{(a_l)^2}+f_{5R}(R,a_l)\dot{R}\dot{a_l}=0
\end{eqnarray}
\begin{eqnarray}\label{eq:18}
\fl
f_{1a_l}(R,a_l)\ddot{R}+f_{2a_l}(R,a_l)\ddot{a_l}+f_{3a_l}(R,a_l)\dot{R^2}+f_{4a_l}(R,a_l)\dot{(a_l)^2}+f_{5a_l}(R,a_l)\dot{R}\dot{a_l}=0
\end{eqnarray}
From this two second order nonlinear differential equation we get
following four first order differential equation, which describes
the growth and growth rate of perturbed interface.
\begin{eqnarray}\label{eq:19}
\frac{d X_1}{d\tau}=X_3
\end{eqnarray}
\begin{eqnarray}\label{eq:20}
\frac{d X_2}{d\tau}=X_4
\end{eqnarray}
\begin{eqnarray}\label{eq:21}
\frac{d X_3}{d\tau}=\frac{n_1d_2-n_2d_1}{m_1n_2-m_2n_1}
\end{eqnarray}
\begin{eqnarray}\label{eq:22}
\frac{d X_4}{d\tau}=\frac{m_2d_1-m_1d_2}{m_1n_2-m_2n_1}
\end{eqnarray}
where, $X_1=\frac{R}{R_2},X_2=\frac{a_l}{R_2}$
$m_1=\xi_1(X_1,X_2),m_2=\xi_2(X_1,X_2),n_1=\xi_3(X_1,X_2)$ and
$n_2=\xi_1(X_1,X_2)$ are nondimensional perturbed radial
displacement, angular displacement, radial velocity and angular
velocity respectively.

\section{Results \& Discussions}
The feasible analytical solution of this set of equations is quite
impossible. We can numerically solve this set of equations using
Runge-Kutta-Fehlberg numerical technique. The numerical results
shows that due to the shock impingement, the interface decreases
and consequently the DT fuel fuel is compressed during
acceleration phase with time. The perturbed interface attains a
minimum position and then kicks back to gradually increasing in
time (Fig.2a) during the deceleration phase. The density increases
as the perturbed interface decreases and attains a maximum density
corresponding to minimum of perturbed interface then the density
gradually decreases (blow-off situation) as the perturbed radius
increases again(Fig.2e). This is the practical situation of
compression-blow off process of ICF. During acceleration phase the
fuel compression is maximum. The dotted line which also represents
the growth of interface with time when $\alpha=\mu R_{2}^3=0$.
However,dash-dot (blue) and solid (red) line for $\alpha=10$ and
15 respectively, represents the stabilization due to density
variation. Our this analytical results show a qualitative good
agreement with simulation${\cite{js05}}$ as well as experimental
results${\cite{lv11}},{\cite{bv11}},{\cite{lv11pl}}$. Though the
experiment has been done in cylindrical geometry, the radius of a
cylinder can be equivalently represented by the radius of a
sphere. We have also shown the polar plot of perturbed surface at
different time (Fig.3,4,5). We have considered  the density ratio
$r=10,{\it l}=50$(Legendre polynomial mode number) for numerical
calculation. This type of density ratio${\cite{km90}}$ is
appropriate for ICF process. We have shown in Fig.6 that the
growth rate is maximum at $l=50$.

\section{Conclusion \& future scope}

Maximum compression can be achieved depending on the density
variation at the interface. The acceleration and deceleration
phase can be better understood from the numerical results.
However, in this work we do not consider the temperature effect
which is more realistic. In future we can take attempt to include
the temperature effect also.

\section{Acknowledgment}
This work is supported by the Department of Science \& Technology,
Government of India under grant no. SR/S2/HEP-007/2008.

{\bf Appendix}

The details of calculation for kinetic energy is given below.
\begin{eqnarray}\label{eq:13}
\fl \nonumber T_{1r\theta}=\frac{\rho_{11}}{2}\int
P_{l}(cos\theta) d\Omega\int ^{\hat r} _{R_1}\Bigg[\frac{R^4\dot
R^2}{ r^2}+2{R^2\dot R}\sqrt{\frac{2{\it
l}+1}{2}}\Bigg(R\dot{a_l}+2\dot{R}a_l\Bigg)\left(\frac{r^{({\it
l}-1)}}{R^{\it l}}\right)P_{l}(cos\theta)\\
\fl \nonumber \hskip 100pt +\Bigg(\frac{2l+1}{2}\Bigg)\Bigg(R\dot
{a_l}+2\dot Ra_l\Bigg)^2\Bigg(\frac{r}{R}\Bigg)^{2 {\it
l}}\Bigg(P_{\it l}(cos\theta)\Bigg)^{2}\\
\fl \hskip 120pt +\frac{1}{{\it l}^2}\Bigg(\frac{2{\it
l}+1}{2}\Bigg)\Bigg(R\dot {a_l}+2\dot
Ra_l\Bigg)^2\Bigg(\frac{r}{R}\Bigg)^{2 {\it l}}\Bigg(\frac{dP_{\it
l}}{d\theta}\Bigg)^{2}\Bigg]e^{\frac{\mu}{3}\Big[R^{3}(t)-r^{3}\Big]}dr\\
\nonumber \fl =\frac{\rho_{11}}{2}\int d\Omega\int ^{\hat r}
_{R_1}\Bigg[\frac{R^4\dot R^2}{ r^2}P_l+2{R^2\dot
R}\sqrt{\frac{2{\it
l}+1}{2}}\Bigg(R\dot{a_l}+2\dot{R}a_l\Bigg)\left(\frac{r^{({\it
l}-1)}}{R^{\it l}}\right)(P_{l})^{2}\\
\fl \nonumber +\Bigg(\frac{2l+1}{2}\Bigg)\Bigg(R\dot {a_l}+2\dot
Ra_l\Bigg)^2\Bigg(\frac{r}{R}\Bigg)^{2 {\it l}}(P_{\it
l})^{3}+\frac{1}{{\it l}^2}\Bigg(\frac{2{\it
l}+1}{2}\Bigg)\Bigg(R\dot {a_l}+2\dot
Ra_l\Bigg)^2\Bigg(\frac{r}{R}\Bigg)^{2 {\it l}}P_{\it
l}\Bigg(\frac{dP_{\it
l}}{d\theta}\Bigg)^{2}\Bigg]e^{\frac{\mu}{3}\Big[R^{3}(t)-r^{3}\Big]}dr\\
\fl=T_{1P_{1}}+T_{1P_{2}}+T_{1P_{3}}+T_{1P_{\theta}}
\end{eqnarray}
Where,
\begin{eqnarray}\label{eq:25}
\fl T_{1P_{1}}=\frac{R^4\dot R^2}{2}\rho_{11}\int P_ld\Omega\int
^{\hat r} _{R_1}\frac{1}{
r^2}e^{\frac{\mu}{3}\Big[R^{3}(t)-r^{3}\Big]}dr
\end{eqnarray}
\begin{eqnarray}\label{eq:26}
\fl T_{1P_{2}}=\frac{\rho_{11}}{2}\int d\Omega\int ^{\hat r}
_{R_1}\Bigg[2{R^2\dot R}\sqrt{\frac{2{\it
l}+1}{2}}\Bigg(R\dot{a_l}+2\dot{R}a_l\Bigg)\left(\frac{r^{({\it
l}-1)}}{R^{\it
l}}\right)(P_{l})^{2}\Bigg]e^{\frac{\mu}{3}\Big[R^{3}(t)-r^{3}\Big]}dr
\end{eqnarray}
\begin{eqnarray}\label{eq:27}
\fl T_{1P_{3}}=\frac{\rho_{11}}{2}\int d\Omega\int ^{\hat r}
_{R_1}\Bigg[\Bigg(\frac{2l+1}{2}\Bigg)\Bigg(R\dot {a_l}+2\dot
Ra_l\Bigg)^2\Bigg(\frac{r}{R}\Bigg)^{2 {\it l}}(P_{\it
l})^{3}\Bigg]e^{\frac{\mu}{3}\Big[R^{3}(t)-r^{3}\Big]}dr
\end{eqnarray}
\begin{eqnarray}\label{eq:26}
\fl T_{1P_{\theta}}=\frac{\rho_{11}}{2}\int d\Omega\int ^{\hat r}
_{R_1}\Bigg[\frac{1}{{\it l}^2}\Bigg(\frac{2{\it
l}+1}{2}\Bigg)\Bigg(R\dot {a_l}+2\dot
Ra_l\Bigg)^2\Bigg(\frac{r}{R}\Bigg)^{2 {\it l}}P_{\it
l}\Bigg(\frac{dP_{\it
l}}{d\theta}\Bigg)^{2}\Bigg]e^{\frac{\mu}{3}\Big[R^{3}(t)-r^{3}\Big]}dr
\end{eqnarray}
After a straight forward calculation of Eq. (25) and putting the
value of $\hat r$ and expanding it upto second order of
$(\frac{a_l}{R})$, we get the following equation
\begin{eqnarray}\label{eq:26}
\fl T_{1P_{1}}=4\pi
R^3\rho_1\Big(\frac{\rho_{11}}{\rho_1}\Big)\Big(\frac{\dot
R^2}{2}\Big)\Bigg[\frac{(a_l/R)}{\sqrt{2(2{\it l}+1)}}\Big(1+\mu
R^3\Big)-\frac{(a_l/R)^2}{2}\Big(1+\frac{\mu R^3}{2}\Big)Q_l\Bigg]
\end{eqnarray}
Similarly,
\begin{eqnarray}\label{eq:26}
\nonumber \fl T_{1P_{2}}=4\pi
R^3\rho_1\Big(\frac{\rho_{11}}{\rho_1}\Big)\Big(\frac{\dot
R^2}{R}\Big)\Big(R\dot {a_l}+2\dot Ra_l\Big)\sqrt\frac{2{\it
l}+1}{2}\Bigg[\frac{({\it l}+3+\mu R^3)}{{\it l}(2{\it l}+1)({\it
l}+3)}+\frac{(a_l/R)}{\sqrt{2(2{\it l}+1)}}Q_l\\
+\Bigg\{\Big({\it l}-1-\mu R^3\Big)\frac{S_l}{4}-\frac{1}{2(2{\it
l}+1)}\Bigg\}\Big(\frac{a_l}{R}\Big)^2\Bigg]
\end{eqnarray}
\begin{eqnarray}\label{eq:27}
\nonumber \fl T_{1P_{3}}=4\pi
R^3\rho_1\Big(\frac{\rho_{11}}{\rho_1}\Big)\Bigg(\frac{2{\it
l}+1}{4}\Bigg)\Bigg(\frac{R\dot {a_l}+2\dot
Ra_l}{R}\Bigg)^2\Bigg[\Bigg(1-\frac{\mu R^3}{2({\it
l}+2)}\Bigg)\frac{Q_l}{(2{\it
l}+1)^2}\\
+\frac{\Big(a_l/R\Big)}{\sqrt{2(2{\it l}+1)}}S_l+\Bigg\{\Big({\it
l}-\frac{\mu R^3}{2}\Big)\frac{X_l}{2}-\frac{1}{(2{\it
l}+1)}\frac{Q_l}{2}\Bigg\}\Bigg(\frac{a_l}{R}\Bigg)^2\Bigg]
\end{eqnarray}
\begin{eqnarray}\label{eq:27}
\nonumber \fl T_{1P_{\theta}}=4\pi
R^3\rho_1\Big(\frac{\rho_{11}}{\rho_1}\Big)\Big(2{\it
l}+1\Big)\Bigg(\frac{R\dot {a_l}+2\dot Ra_l}{2{\it
l}R}\Bigg)^2\Bigg[\frac{{\it l}({\it l}+1)}{(2{\it l}+1)^2({\it
l}+2)}\Big({\it l}+2+\frac{\mu R^3}{2}\Big)T_l\\
+\frac{{\it l}({\it l}+1)}{\sqrt{2(2{\it
l}+1)}}\Big(\frac{a_l}{R}\Big)U_l+\Bigg\{\frac{1}{2}\Big({\it
l}-\frac{\mu R^3}{2}\Big)V_l-\frac{1}{2({2\it l}+1)}T_l\Bigg\}{\it
l}({\it l}+1)\Big(\frac{a_l}{R}\Big)^2\Bigg]
\end{eqnarray}
\begin{eqnarray}\label{eq:15}
\fl \nonumber T_{2r\theta}=\frac{\rho_{21}}{2}\int d\Omega\int
^{R_2}_{\hat r} \Bigg[\frac{R^4\dot R^2}{r^4}+2{R^2\dot
R}\sqrt{\frac{2{\it
l}+1}{2}}\Bigg(R\dot{a_l}+2\dot{R}a_l\Bigg)\left(\frac{R^{{(\it
l}+1)}}{r^{({\it
l}+4)}}\right)\Bigg(P_{l}\Bigg)\\
\fl \nonumber \hskip 100pt +\Bigg(\frac{2l+1}{2}\Bigg)\Bigg(R\dot
{a_l}+2\dot Ra_l\Bigg)^2\Bigg(\frac{R^{2{({\it l}+1)}}}{r^{2({\it
l}+2)}}\Bigg)\Bigg(P_{\it l}\Bigg)^{2}\\
\fl \hskip 50pt+\frac{1}{({\it l}+1)^2}\Bigg(\frac{2{\it
l}+1}{2}\Bigg)\Bigg(R\dot {a_l}+2\dot
Ra_l\Bigg)^2\Bigg(\frac{R^{2{({\it l}+1)}}}{r^{2({\it
l}+2)}}\Bigg)\Bigg(\frac{dP_{\it
l}}{d\theta}\Bigg)^{2}\Bigg]e^{\frac{\mu}{3}\big[R^{3}(t)-r^{3}\big]}r^{2}dr\\
\fl=T_{2P_{1}}+T_{2P_{2}}+T_{2P_{3}}+T_{2P_{\theta}}
\end{eqnarray}
Where,
\begin{eqnarray}\label{eq:27}
\nonumber \fl T_{2P_{\theta}}=4\pi
R^3\rho_2\Big(\frac{\rho_{21}}{\rho_2}\Big)\Bigg(\frac{\dot
R}{2}\Bigg)^2\Bigg[\Big(1+\frac{\mu
R^3}{2}\Big)\frac{(a_l/R)^2}{2}Q_l-\frac{(a_l/R)}{\sqrt{2({\it
l}+1)}}\Big(1+{\mu R^3}\Big)\Bigg]
\end{eqnarray}

\begin{figure}[h]
\begin{center}
\includegraphics[height=2.5in,width=4in,angle=0]{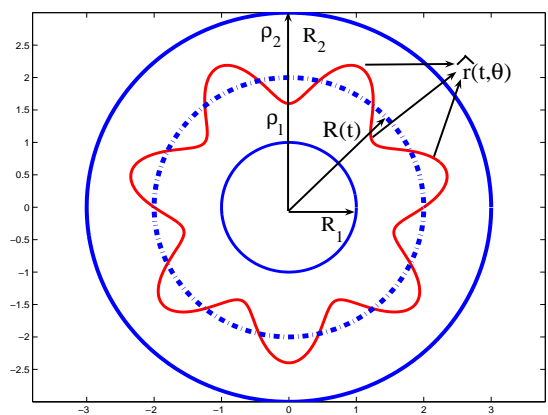}
\caption{Geometry of the problem}
\end{center}
\end{figure}

\begin{figure}[h]
\begin{center}
\includegraphics[height=7.5in,width=7in,angle=-90]{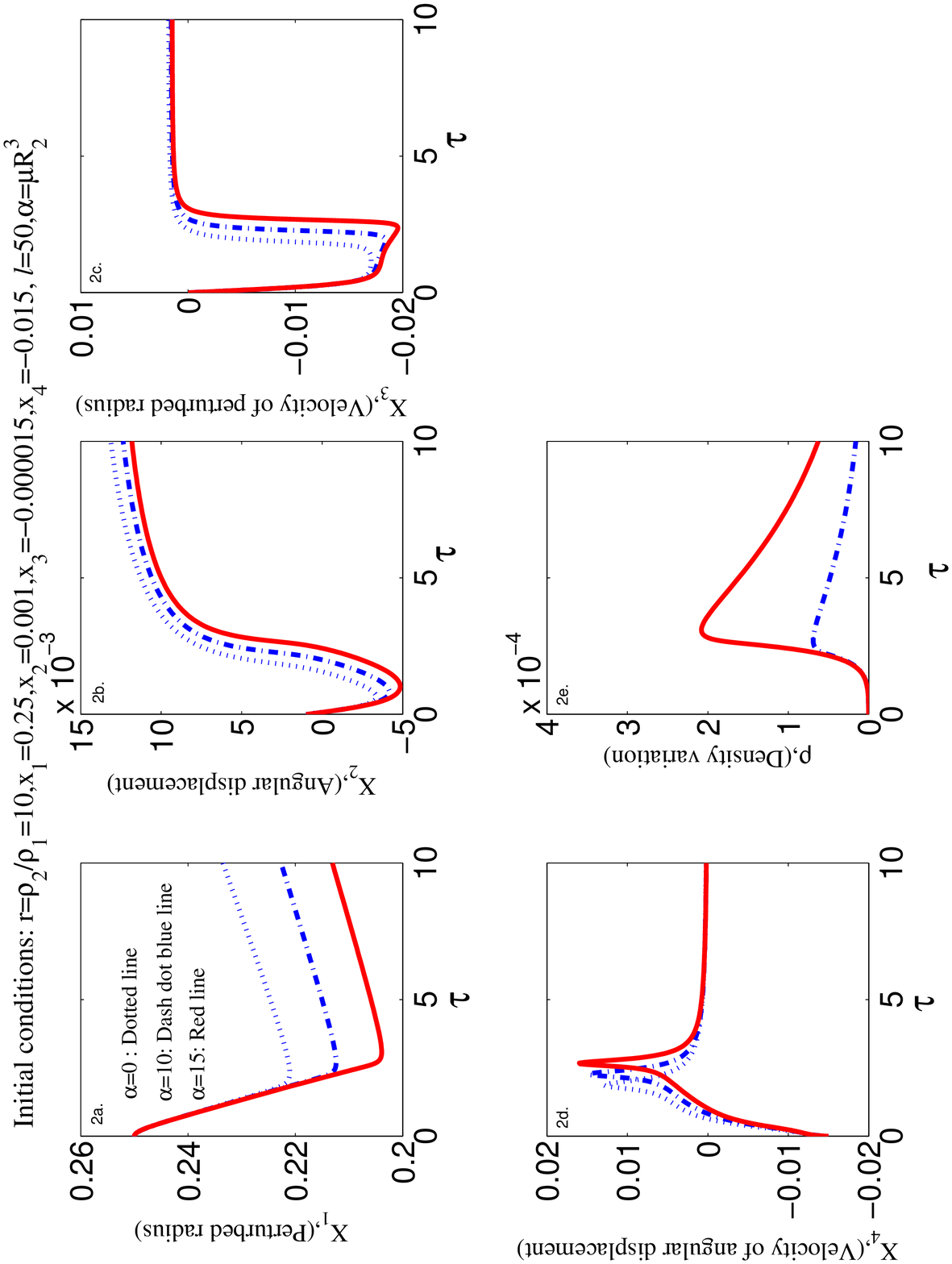}
\caption{}
\end{center}
\end{figure}

\begin{figure}[h]
\begin{center}
\includegraphics[height=4in,width=6in,angle=0]{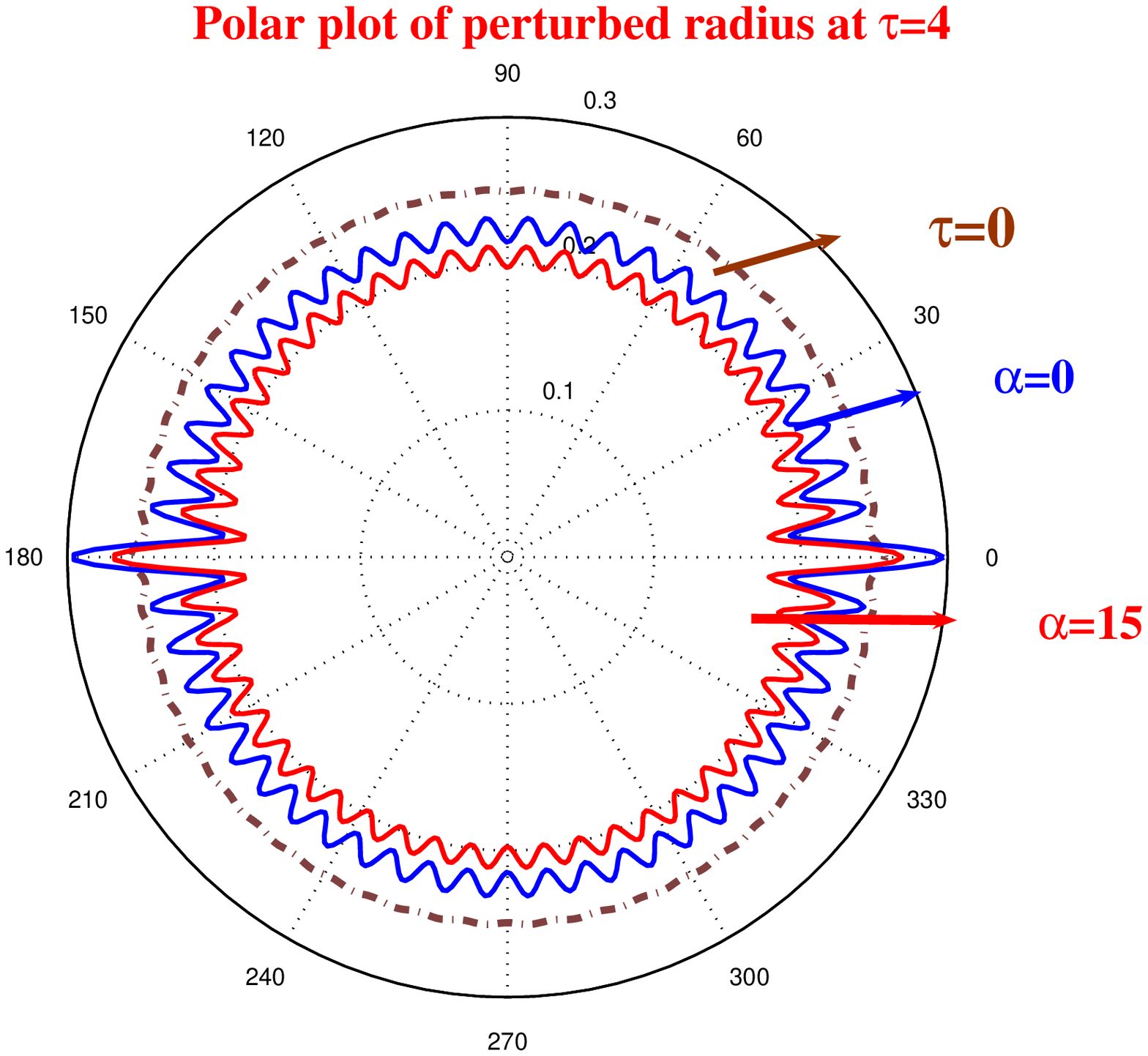}
\caption{}
\end{center}
\end{figure}

\begin{figure}[h]
\begin{center}
\includegraphics[height=4in,width=6in,angle=0]{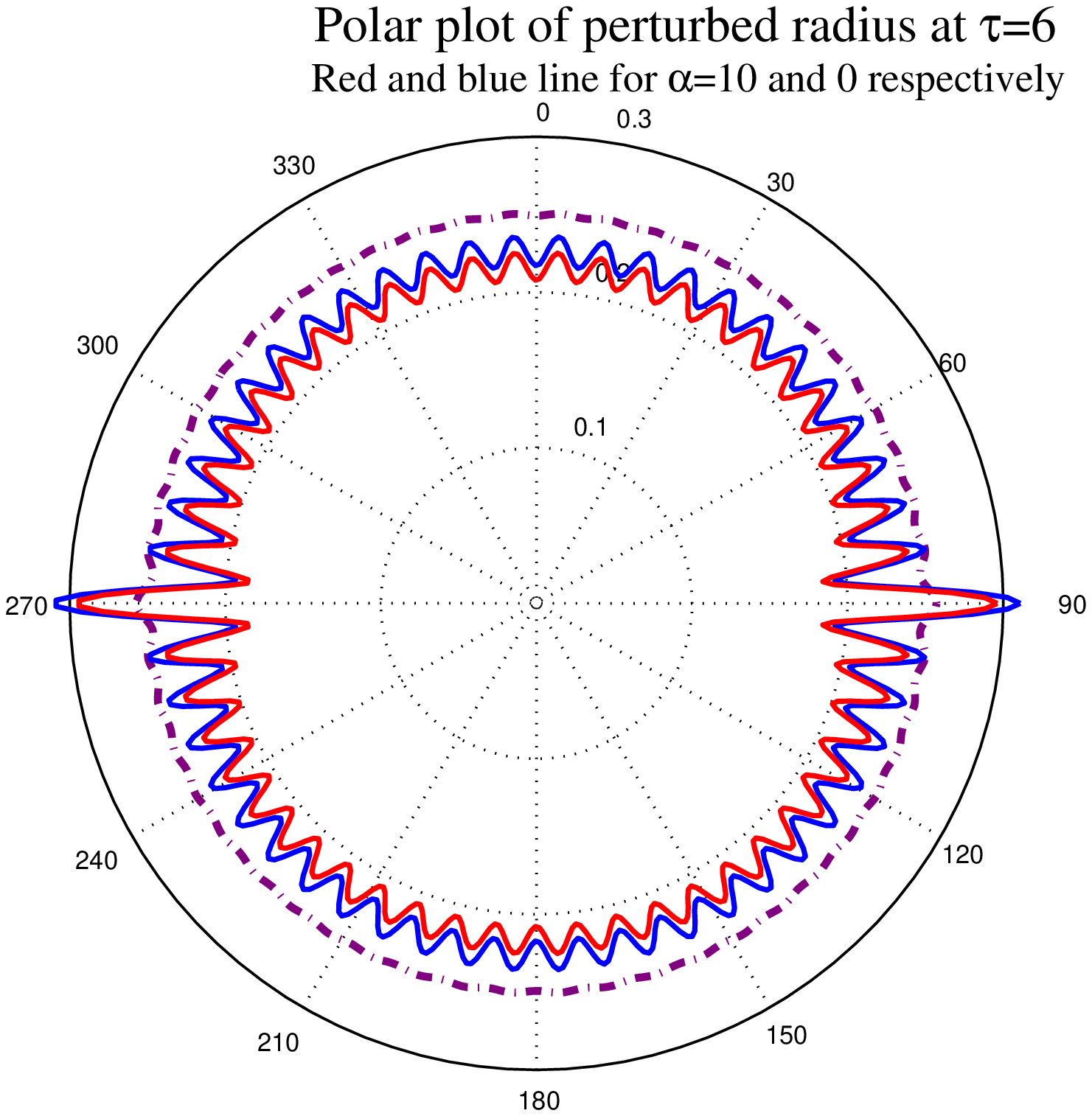}
\caption{}
\end{center}
\end{figure}

\begin{figure}[h]
\begin{center}
\includegraphics[height=4in,width=6in,angle=0]{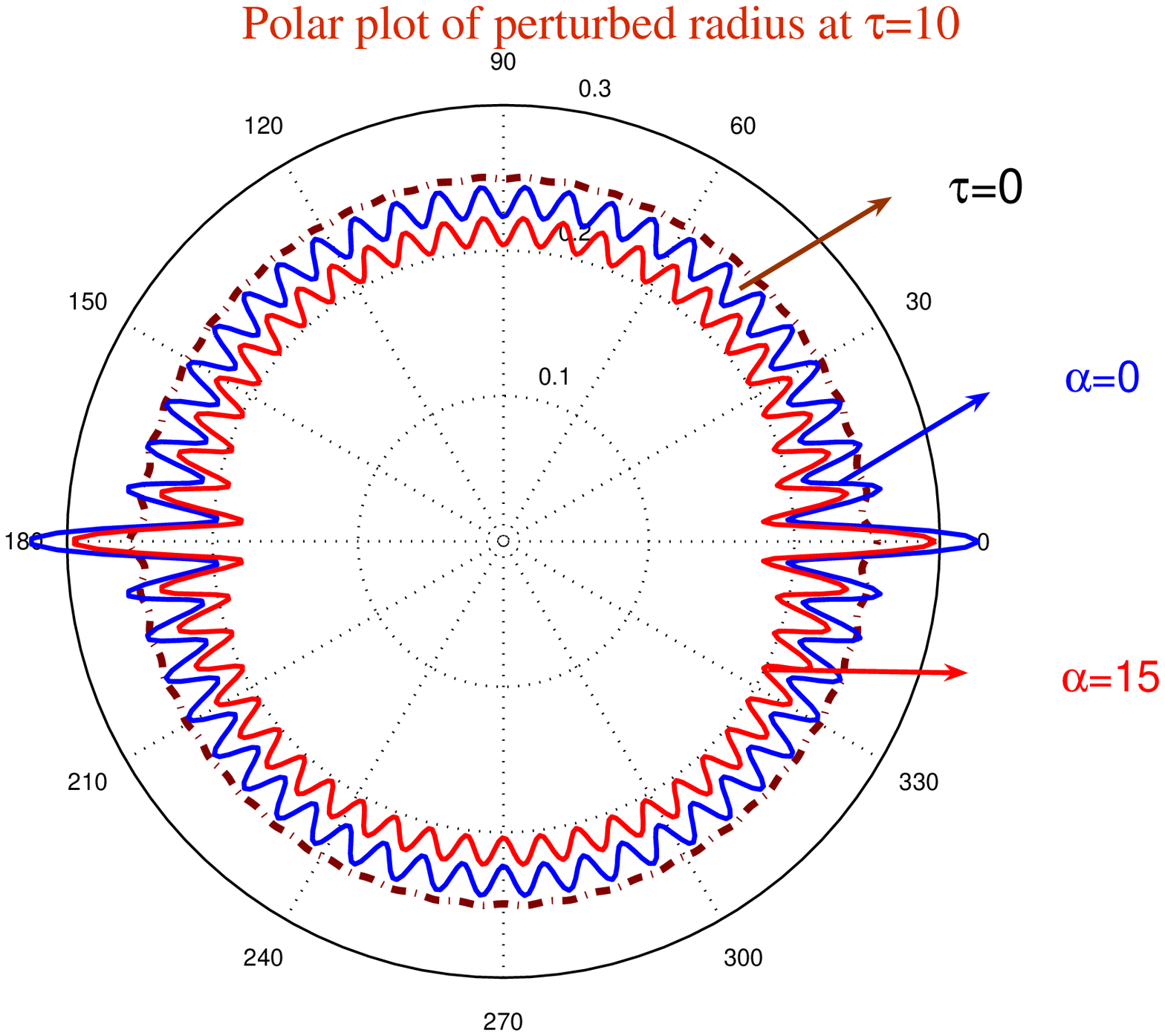}
\caption{}
\end{center}
\end{figure}

\begin{figure}[h]
\begin{center}
\includegraphics[height=4in,width=3in,angle=0]{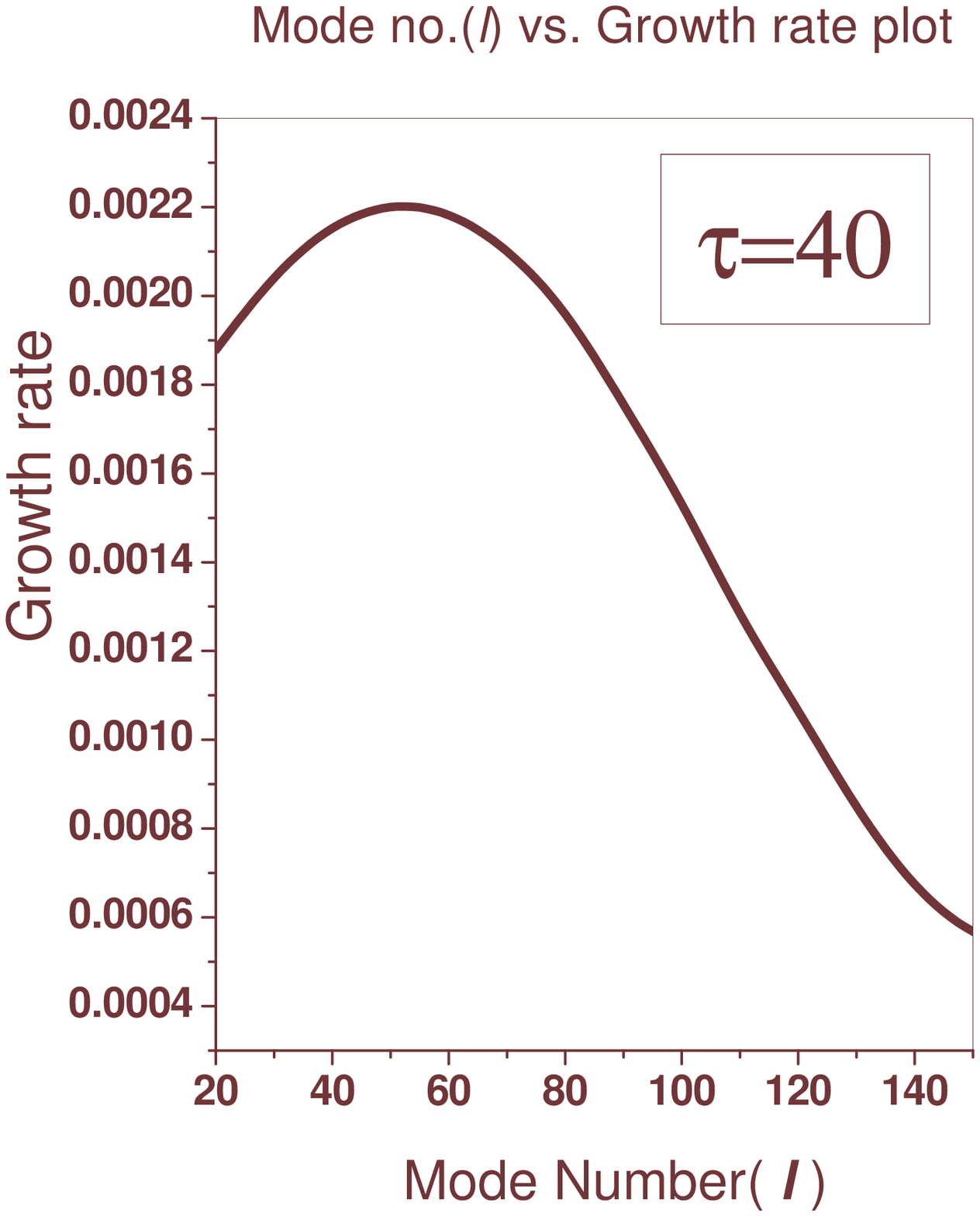}
\caption{}
\end{center}
\end{figure}

\end{document}